\documentclass[floatfix,secnumarabic,amssymb,nobibnotes,nofootinbib,aps,pra,showpacs]{revtex4-2}
\usepackage{epsfig}
\usepackage{amssymb}
\usepackage[colorlinks=true, citecolor=red, linkcolor=blue, urlcolor=blue]{hyperref}
\usepackage{times}
\usepackage{amssymb}
\usepackage{amsmath}
\usepackage{cleveref}
\usepackage{mathrsfs}
\usepackage{graphicx}
\usepackage{epsfig}
\usepackage{dcolumn}
\usepackage{color}
\usepackage{bm}
\usepackage{graphics,psfrag}
\usepackage{graphicx,psfrag}
\usepackage{braket}
\usepackage{soul}
\newcommand{\be}{\begin{equation}}
\newcommand{\ee}{\end{equation}}
\newcommand{\bea}{\begin{eqnarray}}
\newcommand{\eea}{\end{eqnarray}}
\newcommand{\ba}[1]{\begin{array}{#1}}
\newcommand{\ea}{\end{array}}
\renewcommand{\thesection}{\Roman{section}}
\renewcommand{\thesubsection}{\Alph{subsection}}
\usepackage{graphicx,amsmath,amssymb,xcolor,hyperref}
\begin{document}
 \title{Ion-mediated interaction and controlled phase gate operation between two atomic qubits}
 \author{Subhra Mudli$^{\dagger}$, Subhanka Mal$^{\dagger}$, Sinchan Snigdha Rej , Anushree Dey and Bimalendu Deb}
 \email{msbd@iacs.res.in}
 \thanks{\\ \hspace{-0.5in}$^\dagger$These authors contributed equally to this work.}
 \affiliation{School of Physical Sciences, Indian Association for the Cultivation of Science, Jadavpur, Kolkata 700032, India.}
\begin{abstract}
We propose a toy model of ion-atom hybrid quantum system for quantum computing. We show that when two atomic qubits in two largely separated optical tweezers interact with a single trapped ion through Rydberg excitation of the atoms, there exists an ion-mediated atom-atom interaction which exceeds the direct interatomic interaction at large separation. We employ this mediated interaction to demonstrate two-qubit control phase gate operation with 97\% fidelity  by addressing the individual atomic qubits with lasers.
\end{abstract}

\maketitle

\section{Introduction}

A quantum computer works by implementing a universal class of quantum gates with high fidelity. In an ionic or neutral atomic qubit-based quantum computer, single-qubit gate operations are carried out by coherently manipulating  single-qubit states with laser pulses. A two-qubit quantum gate operation requires coherent control of interaction or coupling between the qubits. For high fidelity quantum gates,  the gate operation time must be much smaller than the qubit coherence time. A two-qubit  gate operation \cite{Cirac:PRL:1995,Sorenson:PRL:1999} in an array of largely separated ions in Paul  traps is performed by addressing two ionic qubits individually with laser pulses that control the phononic coupling between the qubits \cite{Leibfried:RMP:2003,Monroe:RMP:2021}. Atoms in  electronic ground   or low lying excited states  generally interact  with a range of  sub-nanometer  scale, ruling out the possibility of generating any micrometer-scale entanglement between such atomic qubits.   An alternative and promising way is to employ long-range ($> 1 \mu$m) interaction between two Rydberg atoms  to accomplish neutral atom two-qubit gate operations. In recent times,  Rydberg atoms in optical lattices or optical tweezers have emerged  as a viable architecture for neutral atom-based quantum computation and quantum simulation \cite{Saffman:RMP:2010,Beterov:JPB:2016,Browaeys:NatPhys:2020,Evered:Nature:2023,Bluvstein:Nature:2024, Basak:PRL:2018}. Based on Rydberg blockade which forbids excitation of a second atom to a Rydberg state when the first atom is already excited,
multi-qubit  quantum gates and programmable quantum algorithm  have been  demonstrated  \cite{Graham:PRL:2019,Graham:Nat:2022,Evered:Nature:2023}.
Rydberg antiblockade which allows two atoms to be simultaneously excited to a Rydberg state is used to construct multi-qubit quantum gates \cite{Su:PRA:2018,Su:PRA:2017}.
With current pace of progress in both neutral atom- and ion-trap technologies, it is expected that in near future a hybrid quantum architecture combining both trapped ions and  atoms will be developed for all or certain tasks in quantum computation and quantum simulation. Of late, ion-atom hybrid quantum systems have attracted a lot of research interests \cite{Tomza:RMP:2019,Niranjan:Atoms:2021,Eberle:JPCS:2015,Jyothi:RSI:2019,Bahrami:arxiv:2023,Cui:arxiv:2023,Ewald:PRL:2019}.

Here we propose a toy model of hybrid quantum system consisting of one trapped ion and two neutral atoms in two separate optical tweezers. We consider that  single atoms in optical tweezers can be brought  in the vicinity of an ion in Paul trap with relatively large separations to avoid any direct atom-ion collision, yet there should be significant ion-atom interaction through Rydberg excitation of the atoms \cite{Secker:PRA:2016,Secker:PRL:2017,Ewald:PRL:2019,Beguin:PRL:2013}. Our  model is schematically shown in Fig.\ref{fig 1} where  two neutral atomic qubits   confined in two similar but largely separate optical tweezers interact  with an ion in a Paul trap.  The center of the ion-trap lies  between the two tweezers' centers. One of the ground-state hyperfine qubit states of the atom may be coupled to  a Rydberg state of the atoms so that the atoms interact with the ion and between themselves primarily through Rydberg excitations. Our proposed hybrid system aims to leverage the advantages of both atom and ion traps,  thereby opening a new perspective in quantum computing.

Here we demonstrate that the utilization of the ion-Rydberg atom interaction  and ion-mediated interaction between two atoms opens a new scope for quantum computation in a hybrid quantum platform. When a ground state of an alkali type   atom  is optically dressed with a highly excited  Rydberg level with principal quantum number $n \sim 100$, the ion-atom  interaction  can be of the order of 10 MHz at a separation of a few tens of microns. Since this interaction at such separations is significantly larger than the typical linewidth $\gamma \sim 10 $ kHz of a Rydberg level, an atomic ground qubit state can be coupled and decoupled with a Rydberg level by lasers at a rate much faster than $\gamma$.
Such strong interaction between an ion and a neutral atom arises due to the $n^7$ scaling of the atomic polarizability. At long separations, the ion-atom interaction potential is of the form $C_4/r^4$ where $C_4$ is a coefficient proportional to the polarizability of the atom due to the electric field produced by the ion. Since induced dipole moment of the atom is the polarizability times the electric field which goes as $1/r^2$, the interaction of this dipole moment with the field leads to the $C_4/r^4$ potential.
The force exerted on the ion by the Rydberg atoms results in a displacement in the ionic position leading to the oscillations of the ionic phonons  which become entangled with the internal states of the atoms.  The Hamiltonian then describes entangled system of two atoms and ionic phonon. Magnus expansion \cite{Magnus:CPAM:1954} of the corresponding evolution operator to the second order in the ratio of the width of the harmonic motional  ground state of the ion to the ion-atom separation  gives rise to an effective Hamiltonian that contains an ion-mediated
Rydberg-Rydberg interaction which exceeds direct atom-atom interaction at large separations. It is interesting to note that this mediated interaction can lead to Rydberg blockade even at a separation on tens of microns, facilitating individual optical addressing of the
atoms within the enlarged blockade radius. In contrast, for the direct Rydberg-Rydberg
interaction, the Rydberg blockade radius is limited to a few micron.
For an alkali-type ground-state atom, the ion-atom interaction at  separations on the order of 10 micron is much less than a Hz, while that becomes quite significant when the atom is excited to a highly excited Rydberg level. Remarkably, the ion-atom interaction at such large separations also exceeds the direct Rydberg-Rydberg interaction.  Typically, the Rydberg-Rydberg interaction at large separations is of van der Waals type while at shorter separations it is of dipole-dipole type \cite{Saffman:RMP:2010}. The van der Waals potential which goes as $1/r^6$ scales as $n^{11}$ \cite{Low:JPB:2012}.  The ion-Rydberg atom interaction will exceed the direct Rydberg-Rydberg van der Waals interaction at a separation larger than a critical value $r_c$ that can be determined by  solving for $r_c$ from the equation $C_4/r_c^{4} = C_6/r_c^{6}$. Since the ion-Rydberg interaction depends on the electric field due to the ionic  charge, this dominating behavior of the ion-atom interaction over the Rydberg-Rydberg direct interaction at separations larger than $r_c$ is universal for different species of ions and Rydberg atoms.
We use this mediated interaction to demonstrate two-qubit controlled phase gate (CZ) operation with about $97\%$ fidelity.
CZ and the controlled-NOT (CX) gates \cite{Nielsen:CUP:2000} are two universal two-qubit gates because any combination of either of these  two-qubit gates with single qubit gates can produce any arbitrary two-qubit entangled states. A CX gate can also be constructed by combination of CZ and single qubit Hadamard gates. So, demonstration of CZ gate suffices to probe that all possible two-qubit gates can be realized in the system under consideration.

Our numerical results with realistic system parameters show that the ion-mediated long-range interaction permits one to realize a two-qubit gate operation between two atomic qubits separated by a distance as large as $21 \mu$m. This means that with our proposed scheme it is possible to entangle two neutral atom qubits at this large separation. To the best of our knowledge, this is the largest distance for two neutral atom qubits to be entangled via controlled interaction. There has been a lot of progress in remote entanglement in linear ion chain in Paul trap \cite{Blatt:Nature:2008} using Coulomb interaction that is always present in the ion-trap platform unlike Rydberg systems.  Remote entanglement has also been achieved in other systems that use heralded entanglement via measurement on the photons \cite{Galli:PRL:2023} and using photons as flying qubits \cite{Krutyanskiy:PRL:2023}. Over the last decade, a number of theoretical and experimental investigations have been carried out in the context of quantum gate operations and entanglement generation using quantum quantum architectures involving superconducting circuit QED \cite{Marcos:PRL:2010, Xiang:PRB:2013, Zou:PRL:2014, Liu:OE:2018, Zhang:OL:2018, Liu:APL:2023, Ma:PRA:2019}.

The paper is organized in the following way. In Sec.\ref{Sec:2} we present  our model, formulate the problem and obtain analytical solutions.
In Sec.\ref{Sec:3} we illustrate the numerical results for a typical system considering experimentally feasible system parameters. Finally we conclude in Sec.\ref{Sec:4}.

\section{ The model and formulation of the problem}\label{Sec:2}
\begin{figure}
\hspace{-0.97in}
  \begin{center}
  \includegraphics[height=2.4in,width=4in]{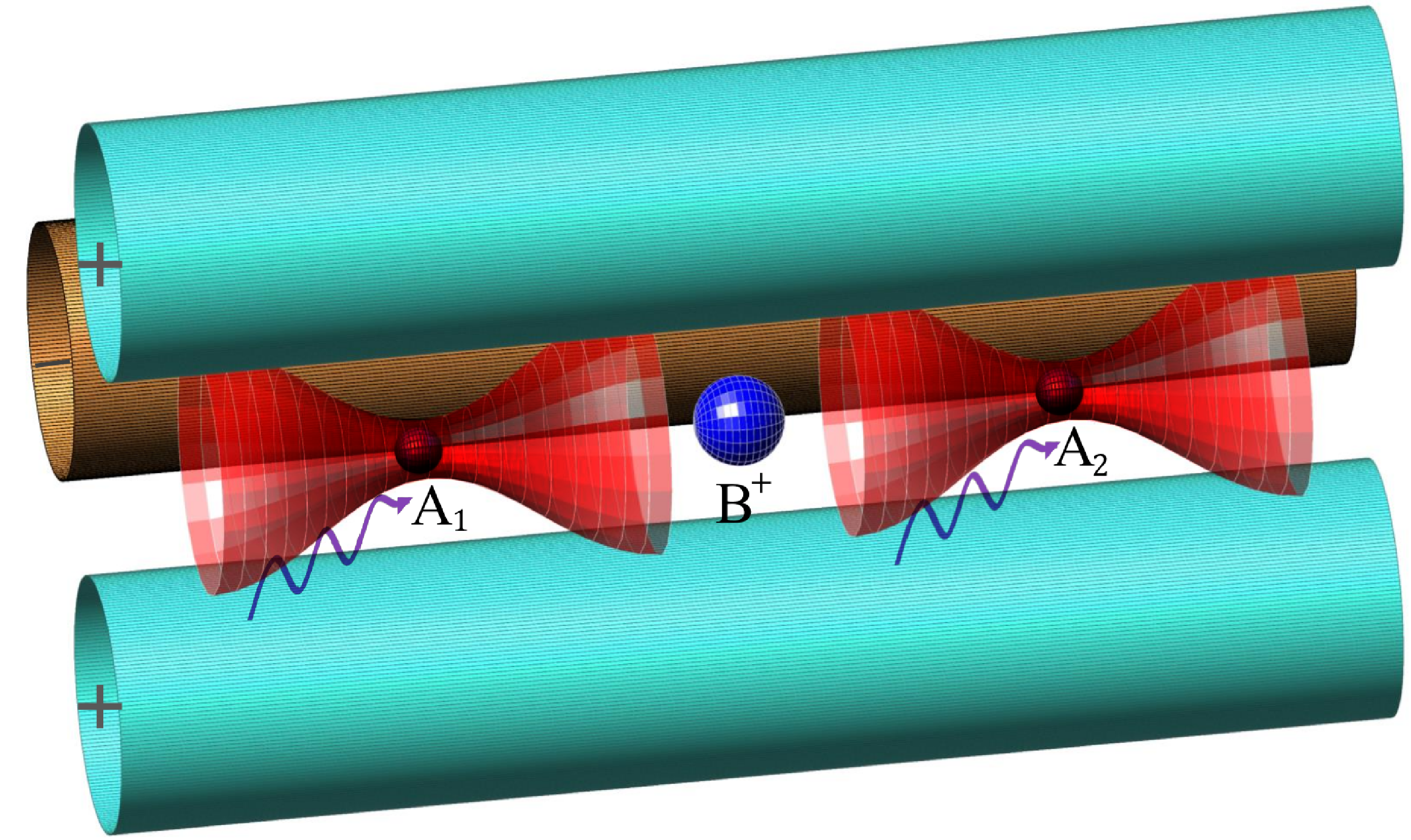}\\
  \end{center}
  \caption{A schematic diagram of the model system where two  atoms $A_{1}$ and $A_{2}$ trapped in two largely separated coaxial optical tweezers are placed on both sides of the ion $B^+$ in a linear Paul trap . The $z$- axis of the ion trap is considered to be collinear with the axes of both the optical tweezers. Both the atoms are trapped in their electronic ground states and are coupled to Rydberg states by separate laser pulses shined on the individual atoms. The interaction of each atom with the ion leads to an effective atom-atom coupling.}
  \label{fig 1}
\end{figure}
A schematic diagram  of our proposed model is shown in Fig.\ref{fig 1}.  Two identical optical tweezers containing identical single atoms $A_{1}$ and $A_{2}$ of mass $m_a$ are placed symmetrically on the two sides of a single ion $B^+$ of mass $m_i$ trapped in a Paul trap. The origin of the coordinate system is set at the electric potential minimum of the ion-trap or the equilibrium position of the ion.

To begin with, let us consider that the atoms have the internal (electronic) states $|0 \rangle \equiv (1 \hspace{0.2cm} 0 )^{T}$, $|1 \rangle \equiv (0 \hspace{0.2cm} 1 )^{T}$ and $|r\rangle$; where $|0 \rangle$ and $|1 \rangle$ denote two ground-state hyperfine levels and $|r \rangle$ a Rydberg state. We  consider that the ground-state sub-levels (hyperfine)  are trapped by optical tweezers and Rydberg states are not trapped. The atomic qubit is composed of  $|0 \rangle$ and  $|1 \rangle$. $| r \rangle$ will be employed as an auxiliary state to couple two qubits for performing two-qubit quantum gate operations.  We assume that  the ion is prepared in the internal ground electronic state $| g \rangle_i$

For simplicity of calculations, we consider an effective one-dimensional (1D) model system along the $z$-direction, assuming the radial trapping frequencies of both the ion trap and the optical tweezers are much higher than the respective axial frequencies and both  the ion  and the two atoms are cooled to the motional ground state of their respective transverse motion.  The ion-atom interaction Hamiltonian $\hat{V}_{\rm{ia}}(z_1, z_2, z_i) = \sum_{j=1,2} \hat{V}_{\rm{ia}}^{(j)}(|z_j-z_i|)$ where
\begin{equation}\label{eq:1}
\hat{V}_{\rm{ia}}^{(j)} =  V_j  |r\rangle_{j} \langle  r| \otimes  | g \rangle_{i} \langle  g |
\end{equation}
Here $V_j = \frac{C_{4}}{|z_j - z_i|^4}$ with $C_{4}$ being the long-range coefficient of interaction between the ion and the atom in the Rydberg state. Here we have ignored the interaction between the ground-state atom and the ion since it is much smaller than that  between a Rydberg atom and the ion at micron scale separation.
As mentioned in the introduction, the  long-range part of the ion-atom interaction, namely, $C_4/r^4$ potential originates from the Coulomb field of the ion that polarizes the atom, and the polarized atom then interacts with the field itself resulting in the said long-range potential. In our model, we assume that two single atoms in optical tweezers are deliberately kept at a sufficiently long distance from the ion  so that the short-range part of the interaction becomes irrelevant and the atoms interact with the ion only through this long-range potential, and thereby avoid any direct collision with the ion.

  Since $|z_j| >\!> |z_i|$, we have
 \bea
 V_j \simeq \frac{C_4}{z_j^4} \left [ 1 + \frac{4 z_i}{z_j} \right ] = V_j^{(0)} + U_j
 \label{eq:2}
 \eea
 where $V_j^{(0)} = C_4/z_j^4$ and
\bea \label{eq:3}
U_j &=& U_j^{(0)} \frac{1}{\sqrt{2}} \left [ a_i e^{- i \omega_i t} + a_i^{\dagger} e^{i \omega_i t} \right ]
\eea
with $U_{j}^{(0)} = V_{j}^{(0)}\beta_{j}$ where $\beta_j = 4 \sqrt{2} \lambda_i/z_j$ is the ratio between width $\lambda_i$  of the motional ground-state probability function  of the ion and the position $z_j$ of the $j$th atom. Here we have written  the position of the ion $z_i$ in the quantized form $z_i =  (1/\sqrt{2})\lambda_i \left [ a_i e^{- i \omega_i t} + a_i^{\dagger} e^{i \omega_i t} \right ] $,   where $\omega_i$ is the harmonic trapping frequency, $\hat{a}_i^{\dagger} (\hat{a}_i )$ represents the creation (annihilation) operator of the 1D harmonic motional quanta (phonon), and $\lambda_i = \sqrt{\frac{\hbar }{ m_i \omega_i}}$ with $m_i$ being the mass of the ion. Since the trapping frequency of the optical tweezers that confine the atomic motion is smaller than that of the ion by two orders of magnitude, we assume that the motion of the atoms is frozen during the time period of the ionic motion and a quantum gate operation time. We therefore do not consider atomic center of mass motion.

\subsection{Ion-mediated interaction}

Consider that the qubit states $|1 \rangle \equiv (0 \hspace{0.2cm} 1 )^{\rm{T}}$ of both the atoms are coupled to the respective  Rydberg state $|r \rangle$ by laser pulses that can individually address the two atoms. The Hamiltonian describing the system is given by $\hat{H} = \sum_{j=1,2} \left [ \hat{V}^{(j)}_{\rm{ia}} + \hat{H}_L^{(j)}\right ] + V_{\rm{r r}} |r r \rangle \langle r r| $ where
\bea
\hat{H}_L^{(j)} &=& - \hbar \delta_j |r \rangle_j \langle r |+
\frac{1}{2}\hbar \left [ \Omega_j |r \rangle_j \langle 1 | + {\rm h.c.}  \right ]
\label{eq:4}
\eea
is the Hamiltonian that describes interaction of the qubits with lasers, $V_{\rm{ r r}} = -\frac{C_{6}}{|z_{1} - z_{2}|^{6}}$ is the direct Rydberg-Rydberg interaction when both the atoms are in Rydberg states and $C_{6}$ is the Van der Waals coefficient for the Rydberg state. Here  $|r r \rangle \equiv |r\rangle_1 |r \rangle_2 $ is the joint state where both atoms are in the Rydberg state $|r \rangle$, $\delta_j = \omega_j - \omega_{rj}$ with $\omega_j$ being the frequency
of the laser that couples the state $|1 \rangle_j$ with the Rydberg state $|r \rangle_j$ and $\omega_{rj}$ the atomic frequency of transition between these two states; $\Omega_j$ is the Rabi frequency for transition $|1 \rangle_j \leftrightarrow |r \rangle_j $.

To derive the ion-mediated interaction we resort to perturbation method considering $\beta_j$ a small parameter since typically $\lambda_i \sim 10 $ nm (nanometer) and $|z_j|$ is chosen to be greater than  1 $\mu$m. Our method relies on the Magnus expansion of the evolution operator $U(t) = \exp[- i \int_0^t \hat{H}(t') d t'/\hbar]$ up to the second order in $\beta_j$. Thus we can write $U(t) \simeq U^{\rm{eff}}(t) = \exp \left [- i \int dt'H^{\rm{eff}}(t') \right] $ where substituting Eq.(\ref{eq:14}) in Eq.(\ref{eq:4}) we  get
\bea
H^{{\rm eff}} &=& \sum_{j=1,2} \left [ \hat{H}_L^{(j)} + V_j^{(0)} (1 + \beta_{j}\hat{\xi}(t) ) |r\rangle_{j} \langle  r| \otimes  |g \rangle_{i} \langle  g| \right ]
+ V_{\rm{ r r}} | r r \rangle \langle r r |
\nonumber \\
&+& \sum_{j=1,2} \hat{\Omega}_j^{\rm{med}} \left ( | r \rangle_j \langle \downarrow | - | \downarrow \rangle_j \langle r | \right ) \otimes | g \rangle_i \langle g | \nonumber \\
&+& V_{\rm{r r}}^{\rm{med}} | r r \rangle \langle r r | \otimes | g \rangle_i \langle g | - \sum_{j=1,2} \frac{ (U_{j}^{(0)})^2}{2\hbar \omega_i} \left [ 1 -  \cos(\omega_i t) \right ] | r \rangle_j \langle r | \otimes | g \rangle_i \langle g |
\label{eq:5}
\eea
where $\hat{\Omega}_j^{\rm{med}}$ is an operator that describes ion-mediated coupling between the state $|r \rangle_j$ and  $|\downarrow\rangle_j$ of the $j$th atom, $\hat{V}_{\rm{r r}}^{\rm{med}}$ is the mediated interaction operator between the two Rydberg atoms. Explicitly, they are given by
\begin{eqnarray}\label{eq:6}
 \hat{\Omega}_j^{\rm{med}} &=& -\frac{iU_{j}^{(0)}\Omega_{j}}{4}\left[ \hat{\xi}(t)t  + \frac{ \hat{\pi}(t)}{\omega_{i}}  + \frac{\hat{\pi}(0)}{\omega_{i}}\right]
\eea
\bea \label{eq:7}
V_{\rm{ r r}}^{\rm{med}} =
 - \frac{U_{1}^{(0)}U_{2}^{(0)}}{\hbar \omega_{i}}\left[1 - \cos{(\omega_{i}t)}\right]
\eea
where
\begin{eqnarray}
\vspace{-0.9in}
 \hat{\xi}(t) = \frac{\hat{a}e^{-i\omega_{i}t}+\hat{a}^{\dagger}e^{i\omega_{i}t}}{\sqrt{2}}\nonumber\\
 \hat{\pi}(t) = \frac{\hat{a}e^{-i\omega_{i}t}-\hat{a}^{\dagger}e^{i\omega_{i}t}}{i\sqrt{2}}\nonumber
\nonumber
\end{eqnarray}
It is to be noted that detailed derivation of $H^{\rm{eff}}$ is given in the Appendix \ref{Appendix-A}.
Note that the mediated interaction results from the second order effect in terms of the small parameter $\beta_j$. Equation (7) shows that in case of equal distance of the two atoms from the ion, the mediated interaction goes as $1/{z_{j}}^{10}$ . However, it is scaled as $n^{14}$ as it is proportional to $C_4^2$ resulting in significant magnitude of the mediated interaction at large separations.
\subsection{Controlled phase gate using ion-mediated interaction}

We use the ion-mediated interaction between the atoms to implement the controlled phase gate operation. This mediated interaction leads to Rydberg blockade even if the atoms are separated by a large distance ($> 10 \mu$m).
CPhase gate implements the transformation
\begin{eqnarray}
 \left\{ |00 \rangle, |01 \rangle, |10 \rangle, |11 \rangle \right\} \rightarrow \left\{|00 \rangle, -|01 \rangle, -|10 \rangle, e^{i\theta}|11 \rangle \right\}
 \label{eq:8}
\end{eqnarray}
where the first atom acts as control and the second one as target. The phase $\theta$ is accumulated on the $|11 \rangle$ state due to the mediated interaction-induced Rydberg blockade. Note that $\theta$ is tunable by changing the Rabi frequency $\Omega$ and gate operation time.

\begin{figure}
\hspace{-0.97in}
  \begin{center}
    \includegraphics[height=2.2in,width=3.7in]{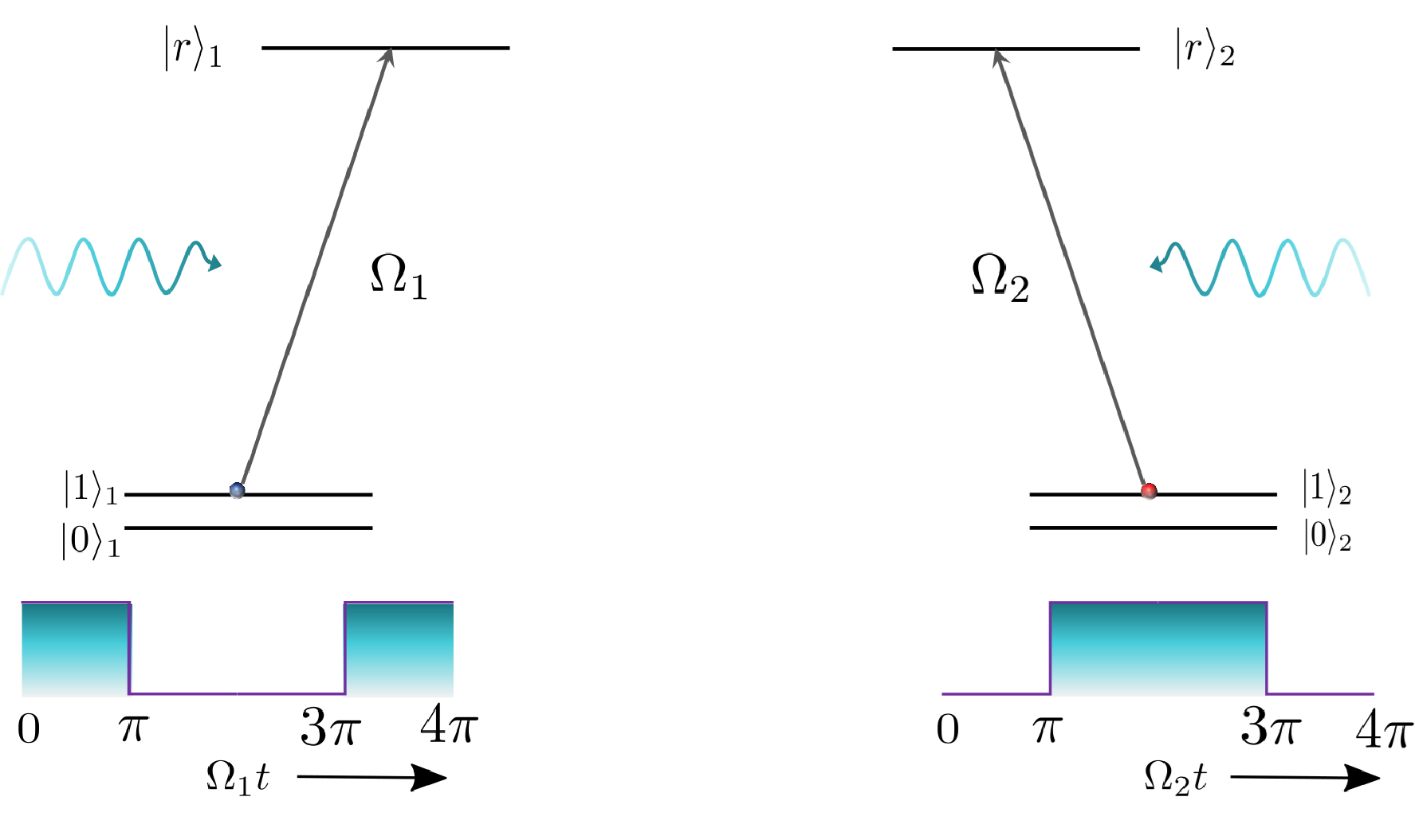}\\
    \includegraphics[height=4.5in,width=4.7in]{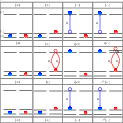}
  \end{center}
  \caption{A schematic diagram for control-phase (CPhase) gate: Here two hyperfine levels in the ground state manifold of an atom constitute the two qubit states $| 0 \rangle$ and $| 1 \rangle$. First $\pi$ pulse acts on control atom, excites it to Rydberg state then a $2\pi$ rotation is given to target atom by applying a $2 \pi$ pulse on this atom, lastly control atom is de-excited to ground state by another $\pi$ pulse. By this pulse sequence a controlled phase gate is realized, with the phase $\theta$ being attributed to the mediated interaction.}
  \label{fig 2}
\end{figure}

Our proposed gate protocol as schematically shown in Fig. \ref{fig 2}, is composed of three consecutive pulses, first pulse is to excite control qubit to a Rydberg state, second pulse is to excite and deexcite target qubit and third pulse is to deexcite the control qubit.
The application of the first $\pi$ pulse on the control qubit is governed by the Hamiltonian
$H_{1}=H_{1}^{\rm{eff}}\otimes\mathbb{I}$. For different initial state preparations one observes the following evolution of states after the first $\pi$ pulse.
\begin{eqnarray}
 |00 \rangle \rightarrow |00 \rangle, |01 \rangle \rightarrow |01 \rangle, |10 \rangle \rightarrow -i|r0 \rangle, |11 \rangle \rightarrow -i|r1 \rangle
 \label{eq:9}
\end{eqnarray}
The Hamiltonian for second pulse is $H_{2}=\mathbb{I}\otimes H_{2}^{eff} + V_{\rm rr}^{tot}| rr\rangle\langle rr|$. This pulse is applied for a duration of $2\pi/\Omega$. After which the states $|10\rangle$, $|01\rangle$ and $|11\rangle$ further evolve to $-i|r0\rangle$, $-|01\rangle$ and $-ie^{i\theta_1}|r1\rangle$, respectively. The effective evolution of the qubits at the end of second pulse therefore becomes
\begin{eqnarray}
 |00\rangle\rightarrow|00\rangle, |01 \rangle \rightarrow -|01 \rangle, |10 \rangle \rightarrow -i|r0 \rangle, |11 \rangle \rightarrow -ie^{i\theta_1}|r1 \rangle
 \label{eq:10}
\end{eqnarray}
Since the second pulse acts only on the second atom coupling  $|1\rangle$  to  $|r\rangle$, only the qubit states of this atom  are modified while the qubit states of the first atom remain unaffected. Finally, a third pulse is applied on the control qubit for a time duration of $\pi/\Omega$ to deexcite the control qubit to $|1 \rangle$ state. The target qubit remains unaffected throughout this operation. As a result the final states become $|00\rangle$, $-|10\rangle$, $-|01\rangle$ and $e^{i\theta}|11\rangle$ from the initial states $|00\rangle$, $|10\rangle$, $|01\rangle$ and $|11\rangle$, respectively. The total phase acquired by the qubit $|11\rangle$ at the end of these pulse sequences is $\theta$.
We can thus realize CZ gate for $\theta=\pi$ implying the gate operator in matrix form
\begin{eqnarray}
U_p = \begin{pmatrix}
1 &0 &0 &0\\
0 &-1 &0 &0\\
0 &0 &-1 &0\\
0 &0 &0 &-1
\end{pmatrix}
\label{eq:11}
\end{eqnarray}

It is worth mentioning that the time-dependent state $| \Psi(t) \rangle$ obtained by operating the evolution operator  $U^{\rm eff}(t)$ on an initially prepared product state  of the two  qubits and the ionic phonon is  a tripartite entangled state involving the internal levels of the two atoms  and the local phonon states of the ion. Explicitly, $| \Psi(t) \rangle$ is given by
\bea
| \Psi(t) \rangle = \sum_{\nu=a_1,a_2,n} C_{\nu}(t) | a_1 a_2 \rangle_a \otimes | n \rangle_i
\label{eq:12}
\eea
where  $ C_{\nu}(t)$ is the probability amplitude, the symbol $\nu$ refers collectively to the atomic and ionic states with $a_1$ and $a_2$ being the atomic levels $0$, $1$ and $r$ of the first and second atom, respectively; and $n$ a number state of the phonon mode. In order to create an entangled state of the two atoms only, one has to make a  projective measurement on a particular phonon number state as we discuss below.

\section{Results and discussions}\label{Sec:3}

For numerical illustration, we consider  {$^{87}$Rb+$^{40}$Ca$^{+}$+$^{87}$Rb} system.  Two hyperfine levels  $|0\rangle \equiv |n=5, L=0, F= 1\rangle$ and $|1\rangle \equiv |n=5, L=0, F= 2\rangle$  of the electronic ground state of  $^{87}$Rb atoms constitute atomic qubit, where $n$, $L$ and $F$ stand for principal, electronic orbital and hyperfine quantum number, respectively. The  level $|1 \rangle$ of both atoms  can be coupled to the Rydberg state $|r \rangle \equiv |n=90, L=0, F=2 \rangle $ by laser pulses as shown in the level diagram of Fig.\ref{fig 3}(a). Since the Rydberg state is considered to be an $S$ ($L=0$) state, this can be coupled to $|1 \rangle$ by a two-photon pulse via a $P$ ($L=1$) state under electromagnetically induced transition scheme \cite{Evered:Nature:2023}. For this Rydberg state $C_{4}$ is $5.07 \times 10^{10}C_{4}^{0}$ \cite{Kamenski:JPB:2014} where $C_{4}^{0} = -160 a.u.$ \cite{Tomza:RMP:2019} is the long-range coefficient of the ion-atom interaction when the atom is in the electronic ground state. The Van-der-Waals coefficient for direct  interaction between two $S$-state $^{87}$Rb Rydberg atoms is $-h\times 16.69$ GHz $\mu$m$^{6}$ \cite{Low:JPB:2012}. Figures \ref{fig 3}(b) and \ref{fig 3}(c) exhibit the variation of $V_{\rm rr}^{\rm med}$ and the direct Rydberg-Rydberg Van der Waals interaction $V_{\rm rr}$ as a function of the interatomic separation $2z_{0}$. For separation smaller than $30.4 \mu$m, $V_{\rm rr}^{\rm med}$ dominates over $V_{\rm rr}$ as Fig.\ref{fig 3}(b) shows.

We consider that the center of each of the optical tweezers is $10.5 \mu$m away from the ion-trap center along the $z$-axis, so that the average separation between the two atoms is $21 \mu$m. For this separation the ion-mediated interaction strength is calculated to be $2\pi\times 0.85$ MHz which is larger than the direct interaction, which is  $2\pi \times 0.19$ MHz at this separation.

 $|V_{\rm rr}^{\rm med}|$ is more than four times larger than $|V_{\rm rr}|$ at the  separation of 21 $\mu$m. However, one may vary the separation to some extent maintaining the ratio $|V_{\rm rr}\rm ^{med}|/|V_{\rm rr}|$ to be large for achieving Rydberg blockade.
Note that $V_{\rm rr}^{\rm med}$ depends on the trapping frequency $\omega_{i}$ of the ion as Eq. (\ref{eq:7}) shows. Since $V_{\rm rr}^{\rm med}$ is second order in $\beta_{j} = 4\sqrt{2}\sqrt{\hbar/(m_{i}\omega_{i}})/z_{j}$, $V_{\rm rr}^{\rm med}$ is proportional to the inverse of $\omega_{i}$. One can vary $\omega_{i}$ over a wide range such that $\beta_{j}$ remains a small parameter. If we fix $z_{j} = 10 \mu$m, then for Ca$^{+}$ ion, for $\beta_{j}$ to become smaller than 0.1, we have $\omega_{i} > 50$ kHz. Typically, ion trapping frequency is of the order of a MHz.

We have set the Rabi frequencies for the transition $|1 \rangle \rightarrow |r \rangle$ for both the atoms at $\Omega = \Omega_{1} = \Omega_{2} = 2\pi \times 0.16$ MHz and the ion trapping frequency at $2\pi \times 0.32$ MHz. We have also considered rate of decay of the Rydberg state to be $\gamma=2\pi\times10$ kHz \cite{Wang:PRX:2022} implying that  the lifetime of the Rydberg  state  is $100 \mu$s. In this parameter regime the mediated interaction is almost $5$ times the Rabi frequency implying Rydberg blockade when the optical transitions  $|1 \rangle \rightarrow |r \rangle$ for the both the atoms are attempted. Here for all the numerical calculations it is considered that $\delta_{j} - \frac{1}{\hbar}V_j^{(0)} = 0$.

\begin{figure}
 \begin{center}
    \includegraphics[height=2.2in,width=2in]{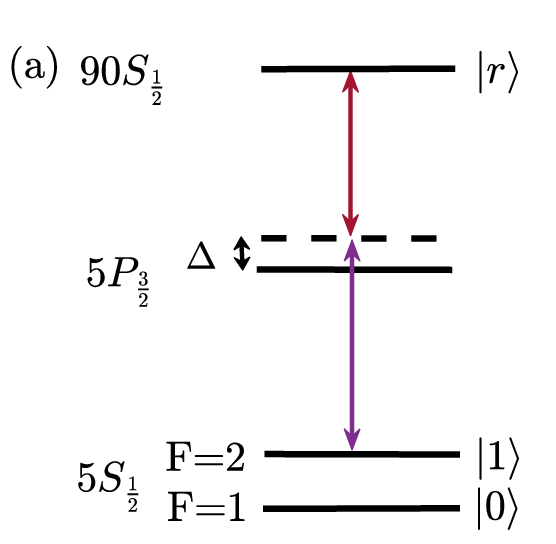}
    \includegraphics[height=2in,width=2.4in]{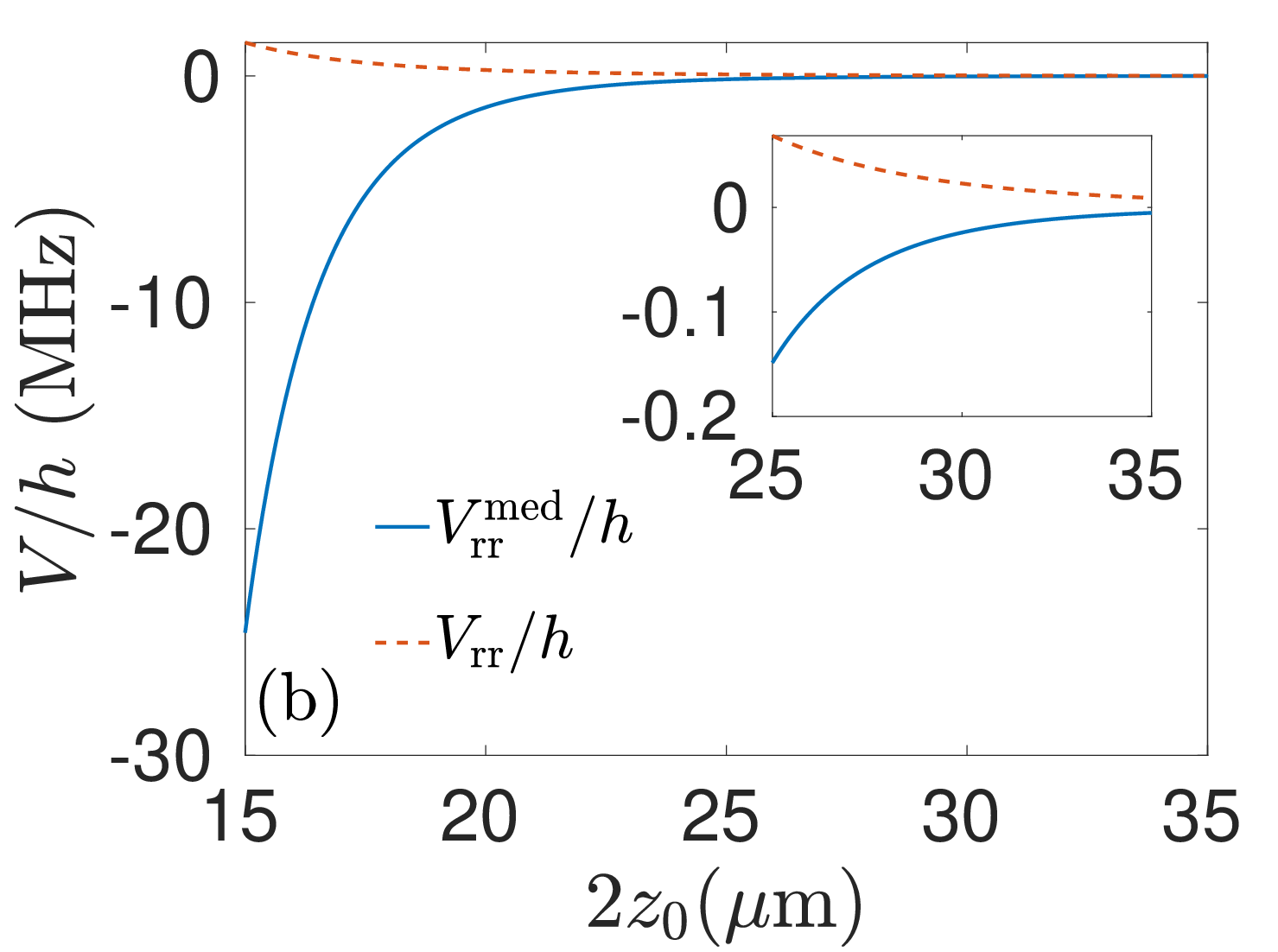}
    \includegraphics[height=2.05in,width=2.4in]{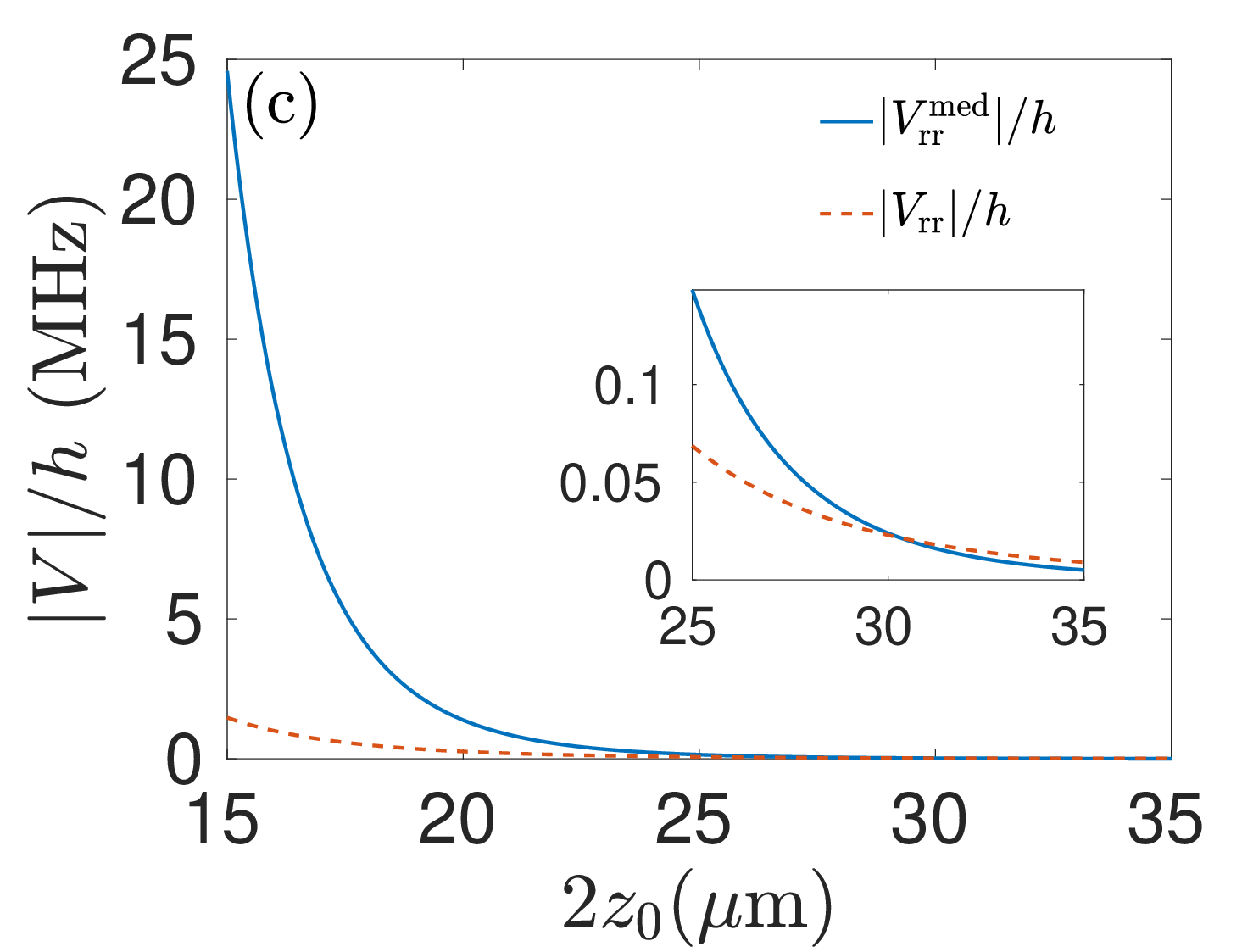}
  \end{center}\vspace{-0.2in}
  \caption{(a) Energy level diagram of a $^{87}$Rb atom showing Rydberg excitation by a two-photon process. The atom is excited from F=2 hyperfine level of $5S_{\frac{1}{2}}$ state ($|1\rangle$) to $90S_{\frac{1}{2}}$ state via the intermediate state $5P_{\frac{3}{2}}$ which is far detuned by $|\Delta|$ under EIT conditions \cite{Evered:Nature:2023}. (b) Ion-mediated Rydberg-Rydberg interaction $V_{\rm rr}^{\rm med}$ (solid) and direct Rydberg-Rydberg intertaction $V_{\rm rr}$ (dashed) are plotted as a function of interatomic separation $2z_{0}$. (c) Same as (b) but for the absolute values of the potentials. Note that  $|V_{\rm rr}^{\rm med}|$ is larger than $|V_{\rm rr}|$ for separation smaller than $30.4 \mu$m for our chosen system.}
  \label{fig 3}
\end{figure}
\begin{figure}
 \begin{center}
    \includegraphics[height=6.4in,width=5.2in]{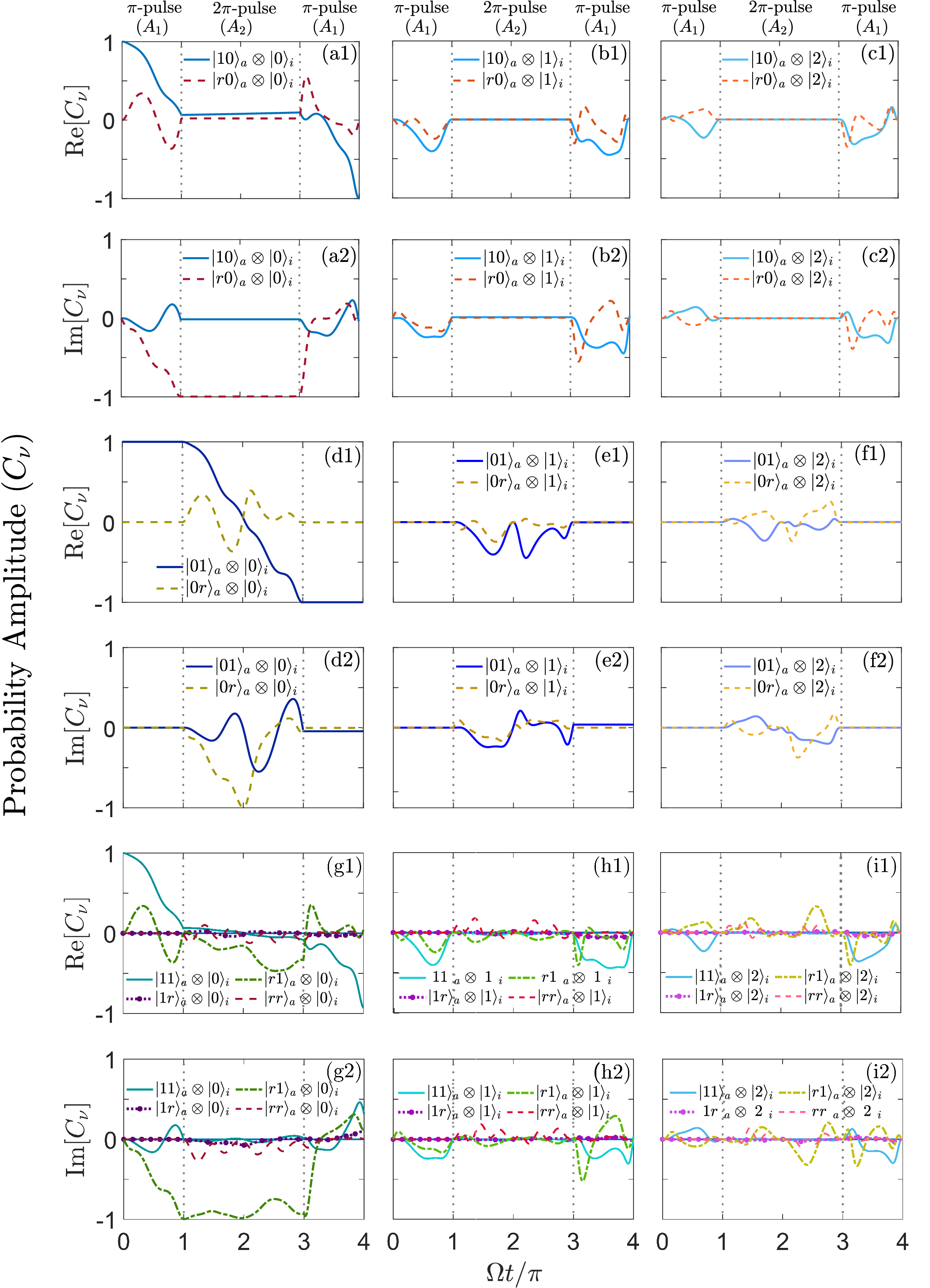}
  \end{center}\vspace{-0.2in}
  \caption{The temporal evolution of probability amplitude ($C_{\nu}$) of different basis states $|\nu \rangle \equiv |a_1 a_2 \rangle_a \otimes |n\rangle_i$ for  different initial joint two-qubit times ionic phonon states. The evolution of real and imaginary parts of $C_{\nu}$  of the  states $|1 0\rangle_{a}\otimes|0\rangle_{i}$ and $|r 0\rangle_{a}\otimes|0\rangle_{i}$ $(\rm {a1, a2})$, $|1 0\rangle_{a}\otimes|1 \rangle_{i}$ and $|r 0\rangle_{a}\otimes|1 \rangle_{i}$ $(\rm {b1, b2})$ and $|1 0\rangle_{a}\otimes|2\rangle_{i}$ and $|r 0\rangle_{a}\otimes|2\rangle_{i}$ $(\rm {c1, c2})$ are shown as a function of dimensionless scaled time $\Omega t/\pi$ with the initial state being $|1 0 \rangle_{a}\otimes|0\rangle_i$. Real and imaginary part of $|0 1\rangle_{a}\otimes|0\rangle_{i}$ and $|0 r\rangle_{a}\otimes|0 \rangle_{i}$ $(\rm {d1, d2})$, $|01\rangle_{a}\otimes|1\rangle_{i}$ and $|0r\rangle_{a}\otimes|1\rangle_{i}$ $(\rm {e1, e2})$ and $|01\rangle_{a}\otimes|2\rangle_{i}$ and $|0r\rangle_{a}\otimes|2\rangle_{i}$ $(\rm {f1, f2})$ states evolve with time when the initial state is $| 01 \rangle_{a}\otimes|0\rangle_i$. The evolution of the states $|11\rangle_{a}\otimes|0\rangle_{i}, |r1\rangle_{a}\otimes|0\rangle_{i}, |1r\rangle_{a}\otimes|0\rangle_{i}$ and $|rr\rangle_{a}\otimes|0\rangle_{i}$ $(\rm {g1, g2})$, $|11\rangle_{a}\otimes|1\rangle_{i}, |r1\rangle_{a}\otimes|1\rangle_{i}, |1r\rangle_{a}\otimes|1\rangle_{i}$ and $|rr\rangle_{a}\otimes|1\rangle_{i}$ $(\rm {h1, h2})$ and $|11\rangle_{a}\otimes|2\rangle_{i}, |r1\rangle_{a}\otimes|2\rangle_{i}, |1r\rangle_{a}\otimes|2\rangle_{i}$ and $|rr\rangle_{a}\otimes|2\rangle_{i}$ $(\rm {i1, i2})$ are also shown with the initial state being $|11 \rangle_{a}\otimes|0\rangle_i$. The separation between two atomic trap centers is $21 \mu$m, $\Omega_{1}= \Omega_{2} = 1$ MHz and decay constant $\gamma = 2\pi\times10$ kHz.}
 \label{fig 4}
\end{figure}
\begin{figure}
\hspace{-0.97in}
  \begin{center}
    \includegraphics[height=3.2in,width=5.1in]{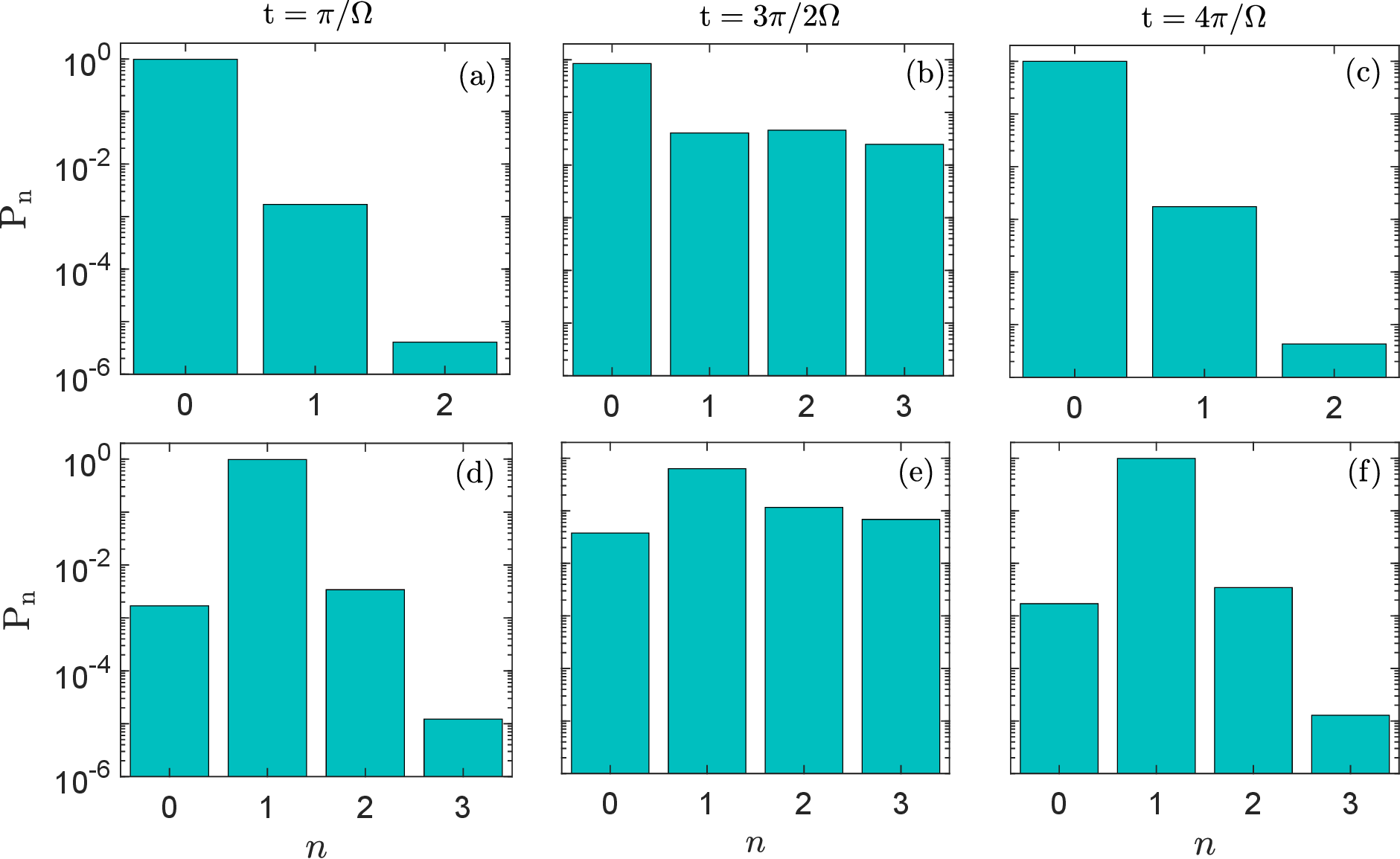}
  \end{center}
  \caption{Phonon distribution ($\rm{P}_{\rm{n}}$) (in semilogarithmic scale) at three different times $t=\pi/\Omega$ (a,d), $t=3\pi/2\Omega$ (b,e) and $t=4\pi/\Omega$ (c,f) for two atomic qubits initially prepared in $|11\rangle_a$ and the ionic phonon prepared in $|0\rangle_i$ (a,b,c) and $|1\rangle_i$ (d,e,f).}
  \label{fig 5}
\end{figure}

When the state is initially prepared in $|1 0\rangle_{a} \otimes |0\rangle_i$,   $|10 \rangle_{a}$  evolves to $-i|r 0\rangle_{a}$ as  Fig.\ref{fig 4}(a1) and (a2) show.  On the other hand, if $|01\rangle_{a} \otimes |0 \rangle_i$ is the initial state, $|01\rangle_{a}\rangle$ is not influenced by the first $\pi$ pulse as can be seen from  Fig.\ref{fig 4}$(\rm{d1})$ and $(\rm{d2})$. In case the state is initially prepared in $|11\rangle_{a} \otimes | 0 \rangle_i$, $|11\rangle_{a}$ evolves to $-i|r1\rangle_{a}$ at the end of the first pulse as shown in Fig.\ref{fig 4}$(\rm{g1})$ and $(\rm{g2})$. The time evolution picture of $|10\rangle_{a}\otimes|0\rangle_i$ in Fig.\ref{fig 4}$(\rm{a1 - c1})$ and $(\rm{a2 - c2})$, $|01\rangle_{a}\otimes|0\rangle_i$ in Fig.\ref{fig 4}$(\rm{d1 - f1})$ and $(\rm{d2 - f2})$ and $|11\rangle_{a}\otimes|0\rangle_i$ in Fig.\ref{fig 4}$(\rm{g1 - i1})$ and $(\rm{g2 - i2})$ reveal that during the action of the second $2\pi$ pulse on the target atom while the control atom is already in the Rydberg state, the target is not excited to the Rydberg state implying the existence of Rydberg blockade due to the mediated interaction. These sub-plots of Fig.\ref{fig 4} also show that after the last $\pi$ pulse on the control atom, the states $| 1 1\rangle_a$, $|10\rangle_a$ and $|01\rangle_a$ acquire a negative sign, while the state $| 0 0 \rangle_a$ remains unchanged. Note that for all these two-qubit states the ionic phonon state is prepared in the ground state $| 0 \rangle_i$. During the time evolution, 1 and 2 phonon states are excited with small probability ($ < 10^{-2}$) while the initial zero phonon state has the largest probability. So, by making a projective measurement on the initial phonon state at the end of all the pulses, one can realize CZ gate between the two atomic qubits which are separated by a long distance.

We have also plotted phonon distribution at some  specific times. We write the density matrix of the entangled state of the two atomic qubits and the ionic phonon as $\rho(t) = |\Psi(t)\rangle \langle\Psi(t)|$. We then obtain the reduced density matrix of phonon by $\rho_{\rm ph}(t) = \rm{Tr}_{\rm a}[\rho]$ where $\rm{Tr}_{\rm a}$ implies tracing over all the atomic states. The phonon distribution is given by $\rm{P_{n}} = \langle n|\rho_{\rm ph}|n\rangle$. In Fig.\ref{fig 5}, we show phonon distribution for different initial conditions. In Fig.\ref{fig 5}(a), (b) and (c) we display phonon distribution at the end of first pulse at $t=\pi/\Omega$, at an intermediate time $t=3\pi/2\Omega$ and at the end of last pulse, i.e, $t=4\pi/\Omega$, respectively, for the state which is initially prepared in $|11\rangle_a$ atomic qubit state and $|0\rangle_i$ ionic phonon state. As mentioned earlier, $P_n$ for $n=1$, $n=2$ and $n=3$ are much smaller as compared to that for  $n=0$ phonon at the end of first and last pulse. In contrast, in the intermediate time, e.g. at $t=3\pi/2\Omega$, the amplitudes of phonon distribution for $n=1$, $2$ and $3$ are not negligible as compared to $n=0$ mode, signifying the Rydberg blockade condition. Figure \ref{fig 5}(d), (e) and (f) show $P_n$ with same initial qubit state $|11\rangle_a$ but with different ionic phonon state $|1\rangle_i$.  For this case, $P_1$  after first and last pulses remains much larger compared to $P_0$, $P_2$ and $P_3$. On the other hand, in the intermediate time at $t=3\pi/2\Omega$, the amplitudes of $P_0$, $P_2$ and $P_3$ are not negligible as compared to $P_1$, similar to the earlier case with $|0\rangle_i$ initial phonon state. Note that the excitation to different phonon states from an initial phonon state depends on the dimensionless coupling parameter $\kappa = U_j^{(0)} \Omega_j/(4\hbar \omega_i)$ Eq.(\ref{eq:6})  between the $j$th atom and the ionic phonon due to Rydberg excitation of the atom. Larger the coupling parameter, larger is the probability for higher number of phonon being involved in the evolution dynamics. In our study, $\kappa < 1$.

From our proposed protocol of CZ gate \cite{Shi:PRAppl:2017, Maller:PRA:2015} we estimate gate fidelity $F$ to be 0.972 by using the formula  $F = \left[| \rm{Tr}(U_{p}U_{i})|^{2} + \rm{Tr}(U_{p}U_{i}U_{i}U_{p})\right]/20 $ \cite{Pedersen:PRA:2007}, where $U_{p}$ is the gate unitary and $U_{i}$ is the unitary of our proposed operation. Fidelity of the state $|11 \rangle_{a}\otimes |0 \rangle_{i}$ is calculated to be 0.983.
In calculating the fidelity, we have considered gate errors due to the spontaneous emission and decoherence of the Rydberg state, and the errors coming out of the coupling of the Rydberg state with multiple ionic phonons. Since the ion is considered to be prepared only in the internal (electronic) ground state, there is no need to consider dissipation or decoherence of the ionic internal states. We have judiciously chosen sufficiently large ion-atom separation so that the ion-Rydberg atom interaction causes small perturbation to the initially prepared phonon state.  Otherwise, excitation of a large number of phonons that remain coupled with the atomic states will lead to dissipation of the qubit states and so degradation of the gate performance. In our calculations, for simplicity we have assumed that the motional state of the ion is initially prepared in 0 or 1 phonon state. However, the ion can be prepared in any arbitrary phonon number state $\mid n \rangle$ as well. But it is necessary to ensure that there is small excitation to nearby phonon states $\mid n+1 \rangle$ and $\mid n -1 \rangle$ while the excitation to other phonon states is negligible. Because, a projective measurement on the initial phonon state is required to decouple the qubit states from the phonon. In our numerical simulation, we have assumed that the lifetime of the Rydberg state is 100 $\mu$s \cite{Wang:PRX:2022} and the time required to perform the gate is 4$\pi$  $\mu$s. Usually, the ionic phonon lifetime is much larger than 100 $\mu$s and so we do not consider phonon dissipation or decoherence in our calculations.

Our proposed ion-atom hybrid quantum platform can offer some unique advantages as compared to other platforms. Usually, for Rydberg atom quantum platforms, it is hard to address the atoms individually within the blockade radius since the laser beam waist is comparable to or larger than the radius. That is why generally atoms in a chain are collectively addressed by shining a laser along the axis of the chain. Then the atoms need to be shuffled from time to time between different sites in order to bring a pair of atoms within the blockade radius and separate them as and when required for various computing tasks \cite{Evered:Nature:2023}. In our proposed hybrid system, the blockade radius is  enhanced far beyond the typical beam size of a laser by leveraging the ion-mediated long-range interaction. Thus our proposed hybrid platform can dispense with the need for atom shuffling. For a linear ion trap, the inter-ion distance is typically of the order of 5 $\mu$m which is maintained by an equilibrium between Coulomb repulsion and strong trapping confinement. In contrast, the distance between the qubits in the proposed hybrid system can be varied over a wide range by choosing different principal quantum numbers of the Rydberg state, yet two qubits remaining within blockade regime and individually addressable. Our hybrid system can be scaled up by adding more neutral atom qubits around the single ion in a two- or three-dimensional configuration.

\section{conclusions and outlook}\label{Sec:4}

In conclusion we have demonstrated a CZ gate with 97\% fidelity by utilizing a trapped ion-mediated interaction between two neutral atom qubits at large separation.   We have considered a toy model consisting of a single ion in a Paul trap and two neutral atoms trapped in two different optical tweezers placed on both sides of the ion in a co-linear geometry. Our calculations with realistic parameters of the system of one Ca$^+$ ion and two $^{87}$Rb atomic qubits have shown that it is possible to  generate an ion-mediated atom-atom interaction when the qubit states are optically coupled to a Rydberg level. Importantly, even at a separation of more than 20 micron the mediated interaction is strong enough to induce a Rydberg blockade that can be utilized for two-qubit quantum gate operation by individually addressing the two qubits with lasers. For numerical illustrations, we have considered an $S$ state of Rydberg level for minimizing ion-trap electric field induced Stark effect on the Rydberg state \cite{Mudli:Arxiv:2023}, however it is possible to consider $P$ or $D$ Rydberg states as well.

In our calculations, we have considered the local phonon due to  ionic motion only. Since the trapping frequency of optical tweezers is much smaller than that of the ion, we have assumed that atomic center-of-mass motion is frozen during the gate operation time. The force between the ion and Rydberg atom causes a small displacement in the ionic equilibrium position leading to the excitation and coherent oscillations of the ionic phonon which in turn become entangled with the atomic internal states. These entangled phonon states may be detected as a measure of quantum sensing of the atomic internal states.

In this study, we have considered only one internal state of the ion. However, it is possible to consider multiple ionic internal states or an ionic qubit that can be coherently manipulated along with the ionic phonon for the purpose of storing quantum information about the atomic qubit states, since ionic qubits  have much longer coherence time ($~ 1$ second). For instance, suppose the ion is initially prepared in the state $| g0 \rangle_i$ with   electronic (internal)  ground state ($g$) and 0 phonon. After the excitation of the control atom to a Rydberg state by a pulse,  there is small probability of one phonon being excited. Now, if a laser field is applied on the ion to affect the transition $|g1 \rangle_i \rightarrow |e0 \rangle_i$ where $e$ refers to a different internal state of the ion,  the two internal states then form a coherent superposition state.  Thus this superposition state can be used as quantum memory for storing the information that the first atom has undergone a transition to a Rydberg state. Next, during the action of the next $2\pi $ pulse on the target atom, there is finite probability of excitation of other nearby phonon states provided the control atom has been excited to the Rydberg state, that is, when the two-qubit state is initially in $|11\rangle_a$ state, as the Fig.\ref{fig 4}(b) and \ref{fig 4}(e) show. Remarkably, the phonon distribution at $t=3\pi/2\Omega$ is markedly different from those at $t=\pi/\Omega$ and $t=4\pi/\Omega$ as a clear signature of Rydberg blockade. thus the information about both qubits initially prepared in $|11\rangle_a$ state undergoing a Rydberg blockade may be inferred from the measurement of phonon distribution \cite{An:NatPhys:2015,Mallweger:PRL:2023,Ohira:PRA:2022}. The information about this phonon distribution may be stored in another internal state of the ion by single-qubit gate operation on the ion. In this way a three-qubit Taffoli gate operation \cite{Nielsen:CUP:2000} may also be possible. Our toy model may be scaled up with multiple atoms and ions to create a hybrid quantum platform for quantum computing and quantum simulation.

There exists surmountable experimental challenges in the implementation of our proposed hybrid system as far as stability of the system is concerned. To overcome these challenges, a number of precautions need to be taken. First,  the optical tweezers should be placed far from the center of the ion trap so that the ion-atom interaction is very small compared to the depth of the optical tweezers' trapping potentials. Since we have assumed that the ground qubit states of the cold atoms are trapped and the un-trapped Rydberg state is excited only for a short duration for which atomic motion can be ignored, it is expected that the system will remain stable for entire duration of the gate operations and beyond. Second, there may be a small Stark shift of the atomic states due to the electric field of the Paul trap electrodes.This shift can be nullified or minimized by assuming that the tweezers’ centers are placed at positions where electric field due to the ion-trap electrodes is small and almost homogeneous. Otherwise, it is possible to consider Stark-shifted Rydberg $S$ level which  remains well isolated from all other Stark-shifted atomic levels at a separation much larger than 0.5 µm \cite{Secker:PRA:2016}.

The ions generally have long coherence time while the Rydberg atoms afford to have strong and long-range interactions mediated through the ion. These features of our hybrid system will be quite useful for implementing quantum computing or creating hybrid quantum technology, for instance, the gate operations can be performed between Rydberg atoms as well as between an atom and ion while ionic qubits can be used as quantum memory.  This single trapped ion surrounded by several atomic qubits in optical tweezers can serve as a basic node for distributed quantum computing or creating quantum network by entangling the atomic qubits with photons which can then be used as flying qubits to carry and distribute entanglement among multiple nodes. This is possible with the currently available technology of ion trapping and optical tweezers. In fact, the difficulties and possible routes to integrating an atom in an optical trap with a single ion in a Paul trap have been already theoretically investigated \cite{Secker:PRA:2016, Ewald:PRL:2019}.

\section{Acknowledgements}

One of us (Subhra Mudli) is grateful to the Department of Science $\&$ Technology, Govt. of India, for DST INSPIRE fellowship.

\appendix
\renewcommand{\thesection}{\Alph{section}} 

\section{} \label{Appendix-A}

Here we derive the effective Hamiltonian of Eq.\eqref{eq:5} using Magnus expansion of
\begin{eqnarray}
U &=& \exp{\left(-\frac{i}{\hbar}\int_{0}^{t}\hat{H}(t')dt'\right)}\nonumber\\ &=& \exp{\left\{-i\left(\hat{\phi}^{(1)}(t) + \hat{\phi}^{(2)}(t) + ...\right)\right\}}
\label{eq:13}
\end{eqnarray}
where $\hat{\phi}^{(1)}(t) = \int_{0}^{t}dt'\hat{H}(t')/\hbar$ and $\hat{\phi}^{(2)}(t) = -\frac{i}{2\hbar^2} \int_{0}^{t}dt'\int_{0}^{t'}dt''[\hat{H}(t'), \hat{H}(t'')]$. Retaining the first two terms in the expansion we can approximate $U(t) = U^{\rm eff}(t) = \exp[-i \int_{0}^{t}\hat{H}^{\rm eff}(t')dt'/\hbar]$. Writing $\hat{H}^{\rm eff}(t) = \hat{H}_{\rm eff}^{(1)}(t) + \hat{H}_{\rm eff}^{(2)}(t)$, where $\hat{H}_{\rm eff}^{(1)}(t) = \hbar \frac{d}{ d t} \hat{\phi}^{(1)}(t)$

\begin{eqnarray}\label{eq:14}
 \hat{H}_{eff}^{(2)}(t) &=&   \hbar\frac{d\hat{\phi^{(2)}}(t)}{dt}\nonumber\\ &=&\sum_{j=1,2}-\frac{iU_{j}^{(0)}\Omega_{j}}{4}\left[\frac{\hat{\pi}(t)}{\omega_{i}} + \hat{\xi}(t)t + \frac{\hat{\pi}(0)}{\omega_{i}}\right] \left(| r\rangle\langle 1| - |1\rangle\langle r|\right)\otimes| g\rangle_{i}\langle g | \nonumber\\ &-&
 \sum_{j=1,2}\frac{(U_{j}^{(0)})^{2}}{2\hbar\omega_{i}}\left[1-\cos(\omega_{i}t)\right] | r\rangle_{j}\langle r| \otimes| g\rangle_{i}\langle g | \nonumber\\ &-&
 \frac{U_{1}^{(0)}U_{2}^{(0)}}{\hbar\omega_{i}}\left[1-\cos(\omega_{i}t)\right] | rr\rangle\langle rr| \otimes| g\rangle_{i}\langle g |
\end{eqnarray}
one obtains the expression of Eq.(\ref{eq:5}).


\end{document}